\documentclass[twocolumn]{aastex62}

\submitjournal{PASP}
\shorttitle{SOWAT: Speckle Observations With Alleviated Turbulence}
\shortauthors{F.~Bosco et al.}

\begin{document}

\title{SOWAT: Speckle Observations With Alleviated Turbulence}

\correspondingauthor{Felix Bosco}
\email{bosco@mpia.de}

\author{F. Bosco}
\altaffiliation{Fellow of the International Max Planck Research School on \\ Astronomy and Cosmic Physics at the University of Heidelberg}
\affiliation{Max-Planck-Institut f{\"u}r Astronomie (MPIA), K{\"o}nigstuhl 17, D-69117 Heidelberg, Germany}

\author{J.-U. Pott}
\affiliation{Max-Planck-Institut f{\"u}r Astronomie (MPIA), K{\"o}nigstuhl 17, D-69117 Heidelberg, Germany}

\author{R.~Sch{\"o}del}
\affiliation{Instituto de Astrof{\'i}sica de Andaluc{\'i}a (IAA-CSIC), Glorieta de la Astronom{\'i}a S/N, E-18008 Granada, Spain}

\begin{abstract}
	Adaptive optics (AO) systems and image reconstruction algorithms are indispensable tools when it comes to high-precision astrometry. In this paper, we analyze the potential of combining both techniques, i.e. by applying image reconstruction on partially AO corrected short exposures. Therefore we simulate speckle clouds with and without AO corrections and create synthetic observations. We apply holographic image reconstruction to the obtained observations and find that (i) the residual wavefronts decorrelate slowlier and to a lower limit when AO systems are used, (ii) the same reference stars yield a better reconstruction, and (iii) using fainter reference stars we achieve a similar image quality. These results suggest that holographic imaging of speckle observations is feasible with $\sim 2-3 \times$ longer integration times and $\sim 3$\,mag fainter reference stars, to obtain diffraction-limited imaging from low-order AO systems that are less restricted in sky-coverage than typical high-order AO systems.
\end{abstract}

\keywords{instrumentation: adaptive optics -- techniques: speckle imaging -- techniques: holographic imaging}

\section{Introduction}
Two successful solutions to atmospheric wavefront perturbations, which reduce the achievable resolution in observational data obtained from ground-based telescopes, are the application of (i) adaptive optics (AO) systems, controlling the pupil plane wavefront, and (ii) image reconstruction techniques.

Today, most instruments mounted to the larger telescopes, with diameters $\gtrsim 4$\,m, make use of AO systems in many different designs to tackle a variety of requirements for the different science goals. As they deliver a good correction over a small field of view (FoV), single-conjugate AO (SCAO) systems are a good choice when the science target is a compact source of few arcseconds size. However, for the studies of extended sources or groups of sources like globular clusters, it is more desirable to achieve a homogeneous correction over the FoV. These are obtained from ground-layer AO (GLAO) systems, being conjugated to the wavefront perturbation in the atmospheric ground layer with several natural or laser guide stars (NGS or LGS), distributed over/ around the FoV.
Examples for such instruments are VLT/HAWK-I in combination with AOF and GRAAL \citep{Pirard04,Casali06,KisslerPatig08,Siebenmorgen11}, and  LBT/LUCI in combination with ARGOS \citep{Rabien10}. \cite{Lu18} study the feasibility of equipping the Keck Observatory with a GLAO system.
The synthesis of both, a multi-conjugate AO (MCAO) system, correcting for low-altitude layers over a large FoV and high-altitude layers in a smaller FoV, is realized for instance in the GeMS system at Gemini-South \citep{Rigaut14,Neichel14} and LINC-NIRVANA at the LBT \citep[e.g.][]{Herbst03,Arcidiacono18}.

Another strategy for obtaining good astrometry and photometry over a large field of view is the application of speckle imaging techniques. Examples like the simple shift-and-add algorithm \citep[SSA,][]{Bates80} or lucky imaging \citep{Fried78,Law06} exploit the nature of speckle clouds in short-exposure observations to reconstruct high image quality by realigning the cloud centroids or discarding clouds with large full width at half maximum (FWHM), respectively. The latter is an expensive technique as more than 90\% of the exposures remain unused.
A more elaborate technique, based on the work of \cite{Primot90}, is to deconvolve point spread functions (PSFs) from the short-exposure images and \cite{Schoedel12} have tested this technique by estimating the instantaneous PSF from bright reference stars in the image data themselves. Recently, \cite{Schoedel13} and \cite{NoguerasLara18} have demonstrated the potential of this \emph{holographic imaging} technique, recovering the diffraction limit over the FoV.

It has been shown that speckle imaging algorithms benefit from working on short-exposure observations from AO-assisted instruments. For instance, the lucky imaging technique benefits as the fraction of \emph{lucky} images increases due to the AO correction, as, e.g., \cite{Velasco16} have tested this on AO-assisted $i'$-Band observations from AOLI \citep[e.g.][]{Velasco18} mounted on the 4.2\,m William Herschel Telescope, and \cite{Law09} have used the LAMP instrument on the 5\,m Palomar Hale telescope to achieve good image cosmetics.

In this paper, we now aim at extending these studies towards the larger 8\,m class telescopes where the holographic image reconstruction is preferred to lucky imaging, as the fraction or probability, $P$, of getting lucky images depends on the telescope diameter $D$ as $P \propto \exp \lbrace - \left(D/r_0\right)^2 \rbrace$ \citep{Fried78}.
Therefore, we simulate speckle clouds with and without AO corrections, and use them to create synthetic observations, see Sect.~\ref{sec:simulations}. These data are analyzed for changes in wavefront decorrelation time scales (Sect.~\ref{sec:atmosphere}), improvements in the expected signal-to-noise ratio (Sect.~\ref{sec:signaltonoise}), and tested in the reconstruction pipeline from \cite{Schoedel13}, see Sect.~\ref{sec:holographicimagereconstruction}. The results are summarized in Sect.~\ref{sec:summary}.

\section{Simulations}
\label{sec:simulations}
\subsection{Point spread functions}
We simulate point spread functions (PSFs) for an 8\,m class telescope with the end-to-end Monte Carlo simulation software {YAO}\footnote{Yorick Adaptive Optics simulation tool (YAO), \url{https://github.com/frigaut/yao}.}, which has been widely used during the development of AO systems \citep[see references in][]{Rigaut13}, for instance for the development of the GRAVITY-CIAO system \citep{Kendrew12}.
We apply a typical Paranal atmosphere structure with the parameters given in Table~\ref{tab:atmosphere} \citep[\emph{nominal case} from Table 2 in][]{Kendrew12} and simulate the performance for a seeing of 1\,arcsec at 500\,nm, which is on the pessimistic end of typical Paranal seeing between $0.8 - 0.9$\,arcsec, see the ESO website\footnote{ESO Paranal observing conditions, \url{http://www.eso.org/gen-fac/pubs/astclim/paranal/seeing/}.}. This corresponds to a Fried parameter of $r_0 = 10.1$\,cm in the optical (32.8\,cm in $H$-band), and a coherence time of $\tau_0 = 0.314 \cdot r_0 / \bar{v} = 4.0$\,ms. We note that YAO implicitly applies Taylor's frozen flow hypothesis\footnote{Taylor hypothesized that the atmospheric perturbations may be approximated by a set of discrete layers, where every layer corresponds to a perturbation pattern constant in time, which is moved across the telescope aperture. The hypothesis has been verified experimentally \citep[e.g.][]{Poyneer09}.} by using discrete atmospheric layers.

\begin{table}
	\centering
	\caption{Parameters of a typical Paranal (discrete) atmospheric layer structure \citep{Kendrew12}.}
	\label{tab:atmosphere}
	\begin{tabular}{crrr}
	\hline \hline
	Layer	& $C_n^2$ fraction (\%)	& Speed (m\,s$^{-1}$)	& Altitude (km) 	\\
	\hline
	1 		& 41					& 10.0			& 0.0			\\
	2		& 16					& 10.0			& 0.3 		\\
	3		& 10					& 6.6				& 0.9			\\
	4		& 9					& 12.0			& 1.8			\\
	5		& 8					& 8.0				& 4.5			\\
	6		& 5					& 34.0			& 7.1			\\
	7		& 4.5					& 23.0			& 11.0		\\
	8		& 3.5					& 22.0			& 12.8		\\
	9		& 2					& 8.0				& 14.5		\\
	10		& 1					& 10.0			& 16.5		\\
	\hline
	\end{tabular}
\end{table}

We define setups for four different kinds of wavefront control during the simulations. These cover an open loop without any correction ("noAO"), a ground layer AO (GLAO) system, a single-conjugated AO (SCAO) system, and an enhanced seeing mode (ESM), where we describe the setups in the following, but see Table~\ref{tab:aodesigns} for characteristic parameters.

\begin{table*}
	\centering
	\caption{Parameters of the AO system simulation setups.}
	\label{tab:aodesigns}
	\begin{tabular}{lccccc}
	\hline \hline
							& noAO		& LBT/ESM   	& \multicolumn{2}{c}{GLAO}			& SCAO 	\\
													\cline{4-5}
	Parameter					&			&			& tip-tilt subsystem 	& LGS subsystem 	& 		\\
	\hline
	loop frequency (Hz)			& 500		& 500		& 500			& 500			& 500	\\
	{guide stars}				& 1 NGS		& 1 NGS	    	& 1 NGS 			& 3 LGS 			& 1 NGS	\\
	-- position					& on-axis		& on-axis		& on-axis			& 20\,arcsec off-axis	& on-axis	\\
	-- brightness (mag)			& 9			& 11			& 8				& 22	(Watt)		& 8		\\
	{wavefront sensors}			& 1 SH		& 1 Zernike     	& 1 SH			& 3 SH			& 1 SH	\\
	-- SH apertures				& 8 			& --		     	& 24 				& 12				& 16 		\\
	-- pixel per subaperture		& 4 			& --		     	& 8 				& 8				& 8 		\\
	-- pixel scale (arcsec)			& 0.3			& --		     	& 0.4				& 0.4				& 0.4		\\
	-- wavelength (nm)			& 500		& 500		& 500			& 500			& 500	\\
	-- read out noise ($e^-$)		& --			& 6			& 3				& 3				& 3		\\
	-- optical troughput (\%)		& --			& 50			& 100			& 100			& 30		\\
	{deformable mirrors}			& 0			& 1 Zernike 	& 1 TT 			& 1 SA		& 1 SA/ 1 TT 	\\
	-- gain					& 0.0			& 0.15		& 1.0				& 0.4				& 0.6/ 0.4	\\
	-- number of {actuators}		& --			& --			& --		& $13\times13$		& $17\times17$/ --	\\
	expected long-exposure FWHM	& seeing	& $0.75\times$ seeing	& \multicolumn{2}{c}{$0.4-0.5\times$ seeing} & diffraction limit	\\
	\hline
	\end{tabular}
	\\\textbf{Notes:} NGS/ LGS: Natural/ laser guide star.
	SA: Stack-array mirror. TT: Tip-tilt mirror.
	All SH-WFSs use the YAO SH method 2.
\end{table*}
The noAO setup is restricted to only measure atmospheric phase perturbations and does therefore not apply any corrections as the gain of the deformable mirror (DM) is set to zero.
The SCAO setup was used as a verification of the simulation setup and is designed such that it produces diffraction limited observations of on-axis science targets.
The template for the GLAO design was the ARGOS system at LBT \citep{Rabien10}, with three laser guide stars (LGSs) for the wide-angle ground layer corrections and a single on-axis natural guide star (NGS) which is serving only for the tip-tilt measurements. The LGSs are placed at 20\,arcsec radial distance from the science target, where one is set in the west and the other two are regularly placed at an azimuth angular distance of 120$^\circ$ from the first around the science target. Using LBT/LUCI with the ARGOS system reduces the FWHM of the seeing disk by a factor of $2-2.5$.
The ESM is another observing mode offered for LBT/LUCI which by design is similar to a SCAO system but restricted to correct only for the Zernike orders $\leq11$. \cite{Rothberg18} describe this mode and report that this correction already reaches a reduction of the seeing disk FWHM to $0.5\times$ the natural value, where we adopted the conservative value of 0.75 for our simulations, based on the details on the AO modes offered for observations with LUCI, see the website\footnote{LBT/LUCI observing modes, \url{https://sites.google.com/a/lbto.org/luci/preparing-to-observe/ao-esm-and-argos}}. In their report, they also mention that the PSF is fairly homogeneous up to 2.5\,arcmin away from the reference star.

We run every setup using the same atmospheric structure and the same YAO-phase screens. The iteration time of the simulation is set to 2\,ms, corresponding to a AO-loop frequency of 500\,Hz, where the first ten iterations each are neglected to allow the system to settle. The simulations cover a time interval of 20\,s, where we only used the full variety of PSFs of the noAO and GLAO simulations to obtain a larger variety of short-exposure PSFs for the generation of synthetic observations below. The final setups where tuned to fulfill the expected long-exposure FWHM values from Table~\ref{tab:aodesigns}.

\subsection{Synthetic observations}
We generate synthetic observations by reproducing the imaging process with the \textsc{python} package \textsc{vegapy}\footnote{VEGAPy: A Virtual Exposure Generator for Astronomy in Python, \url{https://github.com/felixbosco/vegapy}}, which is utilizing extensively the \textsc{astropy} package \citep{Astropy13}. The code divides the procedure into three domains: (i) The science target domain, (ii) the (telescope) optics domain, and (iii) the detector domain. We describe the three domains in the following.

\begin{figure}
	\centering
	\includegraphics[width=.45\textwidth]{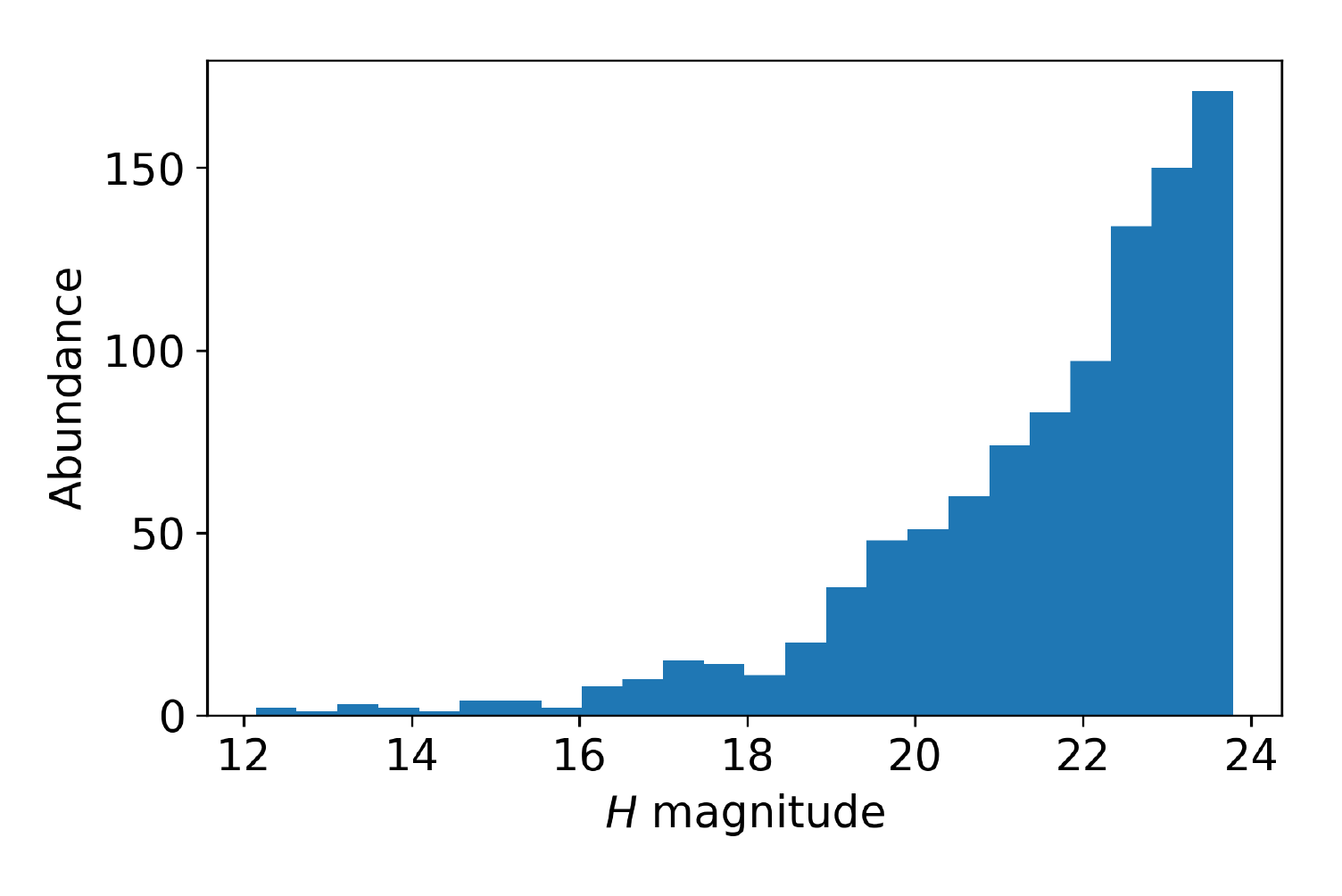}
	\caption{Histogram of the number of stars per $H$-band magnitude bin of the simulated stellar cluster with 1000 objects.}
	\label{fig:histogrammagnitudes}
\end{figure}

The science target is a static image object in units of photons\,m$^{-2}$\,s$^{-1}$, containing the stellar flux values and the sky background flux, e.g., for an $H$-band night sky of 14.4\,mag\,arcsec$^{-2}$ \citep{Cuby00} for a given FoV of $21.6\times21.6$\,arcsec. The magnitudes were converted into flux with the band-specific reference flux values from \cite{Campins85}, giving photometric zero points for Vega of about 1600, 1080, and 670\,Jy in the JHK bands, respectively. We create images for two types of stellar systems, (i) a stellar cluster with a distribution of $H$-band magnitudes as presented in Fig.~\ref{fig:histogrammagnitudes}, distributed randomly across the FoV, for the tests in the reconstruction pipeline, and (ii) a regular grid of stars with well known magnitudes for the signal-to-noise ratio (SNR) measurements.

In the domain of the (telescope) optics, we multiply the science target image by the telescope collecting area and convolve the result with a normalized PSF from above. This may be either a long-exposure PSF or, in the case of short exposures, we integrate the short exposure PSFs to the required discrete integration time (DIT). We note that we do not consider anisoplanatic effects, in particular variations of the PSF across the FoV, in this work. The study of these shall be addressed with observational data, in a folow-up publication. The throughput of the optics is assumed to be on the order of 90\%.

During the following steps in the detector domain, we will include the effect of photon noise by using a copy of this image, being filled with Poisson-distributed random numbers with the original image value as expectation value. After resampling this onto the detector grid, by also considering the detector FoV, we convert the photon number to a number of electrons with the detector quantum efficiency. We add detector specific read-out noise (RON) electrons by adding normal distributed random numbers with a standard deviation corresponding to a literature value of the simulated detector type. In the end, we convert the resulting image to ADU using the detector gain. The applied detector parameters follow the example of a Teledyne HAWAII-2RG detector \citep{Loose07}, windowed down to a $1024\times1024$ pixel detector grid, where the pixel scale is 0.0106\,arcsec, a tenth of the VLT/HAWK-I instrument pixel size \citep[e.g.][]{Pirard04}, corresponding to a detector FoV of $10.9 \times 10.9$\,arcsec. Such detectors have quantum efficiency of 90\%, or 0.9 electrons per photon, and a read-out noise of $\sim35$ electrons per pixel for a fast single-read read-out mode. We set the gain to 17 electrons per ADU to obtain a read-out noise of $\sim2$\,ADU.

We do not consider effects of dark current as the number of electrons due to this effect is expected to be negligible for the short exposure times of order 1\,s relevant for SOWAT observations.

\section{Wavefront decorrelation in the simulated atmosphere}
\label{sec:atmosphere}
For speckle imaging techniques, it is crucial that the exposure times are sufficiently short such that the atmospheric turbulence may be treated as frozen. The time scale for this is the atmospheric coherence time, $\tau_0$. However, \cite{Schoedel13} have shown that holographic imaging works as well for integration times up to $ \gg 10\times\tau_0$, at the cost of a lower Strehl ratio in the resulting reconstruction. In this section, we compare the decorrelation of the atmosphere for the four setups of wavefront control.
We study this behavior in our simulation data on the residual wavefront and derive expectation values for the time scale of the wavefront decorrelation.

\subsection{Instantaneous residual wavefront rms}
In a first step, we analyze the root mean square (rms) of the instantaneous residual wavefronts and compare the results to the corresponding values predicted by \cite{Noll76}. He derives the residual mean square error of a corrected wavefront, i.e. the residual phase variance, where $\Delta_J$ is the residual phase variance after correcting for the first $J$ Zernike modes:
\begin{eqnarray}
	\Delta_1 &= 1.0299 (D / r_0)^{5/3} \, \mathrm{rad}^2 = \Delta_\mathrm{piston} \\
	\Delta_3 &= 0.134 (D / r_0)^{5/3} \, \mathrm{rad}^2 = \Delta_\mathrm{tip-tilt} \\
    \Delta_{11} &= 0.0377 (D / r_0)^{5/3} \, \mathrm{rad}^2 = \Delta_\mathrm{ESM}
	\label{eq:wavefrontvariance}
\end{eqnarray}
In this notation, $\Delta_1$ corresponds to the piston-removed wavefront error (WFE) and $\Delta_3$ is the corresponding WFE after removing the tip-tilt. \cite{Noll76} notes that the phase variance over finite apertures is infinite for a Kolmogorov spectrum, $\Delta_0 = \infty$, whereas this quantity becomes finite after correcting for the piston variance. The YAO output wavefront data are already subtracted by the piston contribution and we compare our results to $\Delta_1$. For an 8.2\,m telescope and $r_0 = 10.1$\,cm in the optical, corresponding to a seeing of 1\,arcsec, we expect rms values of the noAO residual wavefronts in the optical and $H$-band to be 39.7\,rad and 12.2\,rad, respectively.

\begin{figure*}
	\centering
	\includegraphics[width=.95\textwidth]{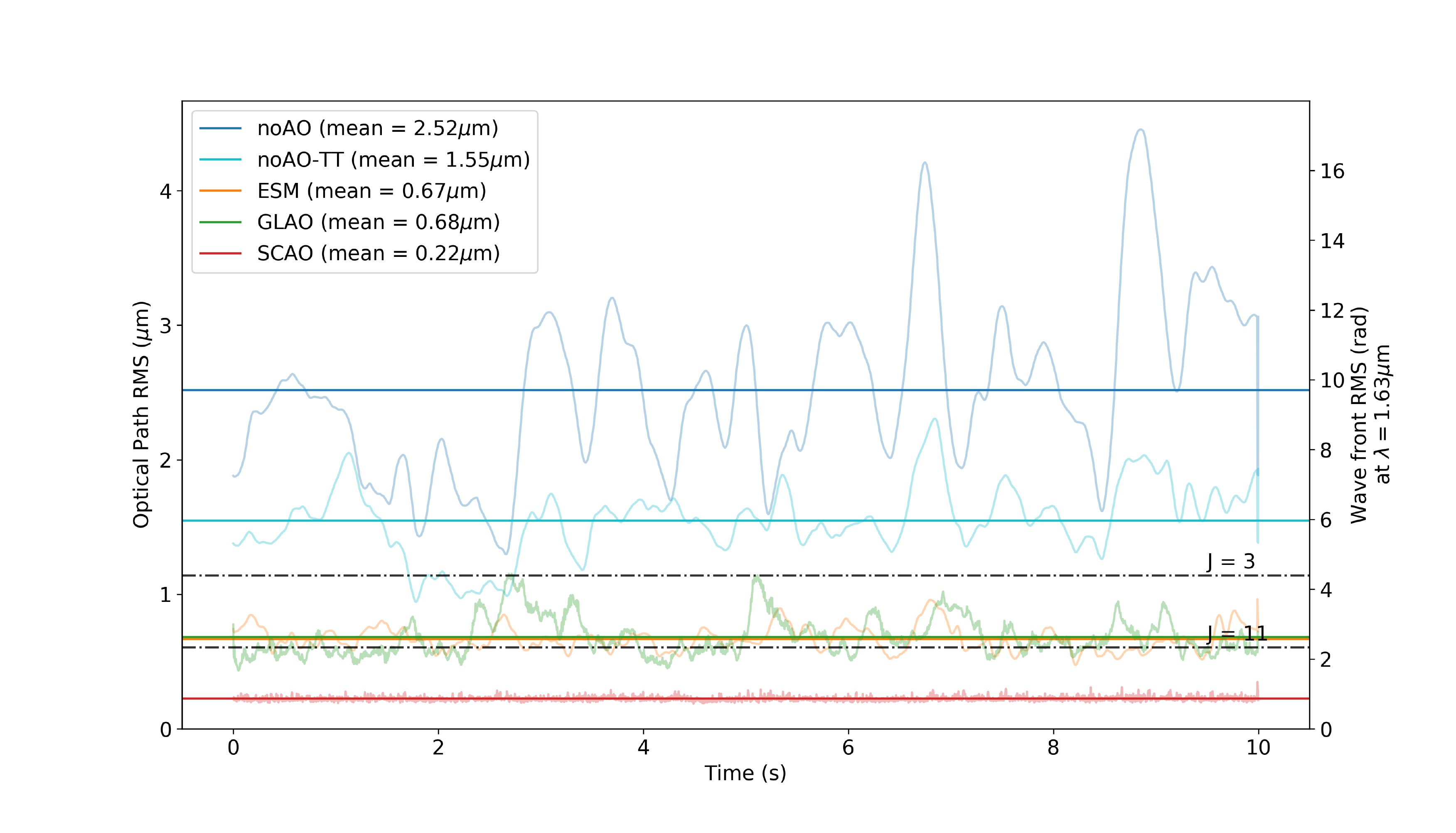}
	\caption{Aperture rms of the residual wavefronts as a function of time. The horizontal lines indicate the mean value of the corresponding series. The dash-dotted black lines indicate the expected values for the atmospheric wavefront rms after correcting for the Zernike orders with given Noll index $J$, based on the results of \cite{Noll76}. $J= 3, 11$ correspond to a correction of tip-tilt, and spherical aberrations, respectively. In the text, we discuss the slight discrepancy between the nominally matching modes, i.e.  $J= 3$ -- noAO-TT, and $J= 11$ -- ESM.}
	\label{fig:wavefronttrends}
\end{figure*}

In Fig.~\ref{fig:wavefronttrends}, we present the rms of the instantaneous residual wavefronts as a function of time for the four simulation setups. We add an additional curve for the tip-tilt correction, by subtracting a least-squares fitted plane from the noAO data.
Besides the pure noAO curve (\emph{blue}), all curve means are (slightly) above the corresponding Noll predictions.
The deviation in the noAO case is expected to be due to the implicit outer scale in the YAO phase screens (M. van Dam, \emph{private communication}).
As we match our long-exposure PSFs to a given seeing level -- while implicitly applying reduced power in the low spatial frequency phase aberrations with respect to the Kolmogorov spectrum -- our AO simulations under-compensate the higher spatial frequency phase aberrations (as visible in the Fig.~\ref{fig:wavefronttrends}).
However, since all the results are derived from simulations with the same atmosphere, especially with the same phase screens, we do not expect a loss of generality in the results.

\subsection{Wavefront decorrelation}
\label{sec:wavefrontdecorrelation}
In a second step, we analyze the wavefront decorrelation time scale for the noAO, ESM and GLAO setups. Therefore, we compute the mean rms of all available wavefront differentials $\Delta\phi(\Delta t) = \phi(t) - \phi(t + \Delta t)$ for a number of time intervals $\Delta t$, where we average over all $t$ with $t + \Delta t \leq 10$\,s:
\begin{equation}
	\mathrm{rms} \left\lbrace \Delta\phi(\Delta t) \right\rbrace \equiv
    \left\langle \mathrm{rms} \left\lbrace \phi(t) - \phi(t + \Delta t) \right\rbrace \right\rangle
\end{equation}
This quantity is expected to grow time-wise as the two snap-shots of the atmosphere are statistically increasingly uncorrelated. As a result of the finite outer scale $L_0$ of atmospheric turbulence, however, this trend does not continue until infinity but the variance converges to a maximum achievable phase variance \citep[cf. Fig.~4.4 in][]{Glindemann11}. Therefore, we model this trend with a bounded exponential growth starting from $f(0)=0$, with a boundary $B$ and growth constant $k$:
\begin{equation}
	f_{k, B}(t) = B \cdot \left(1 - \exp\left\lbrace - k t \right\rbrace \right)
    \label{eq:bounded_growth_fit}
\end{equation}

\begin{figure*}
	\centering
	\includegraphics[width=.95\textwidth]{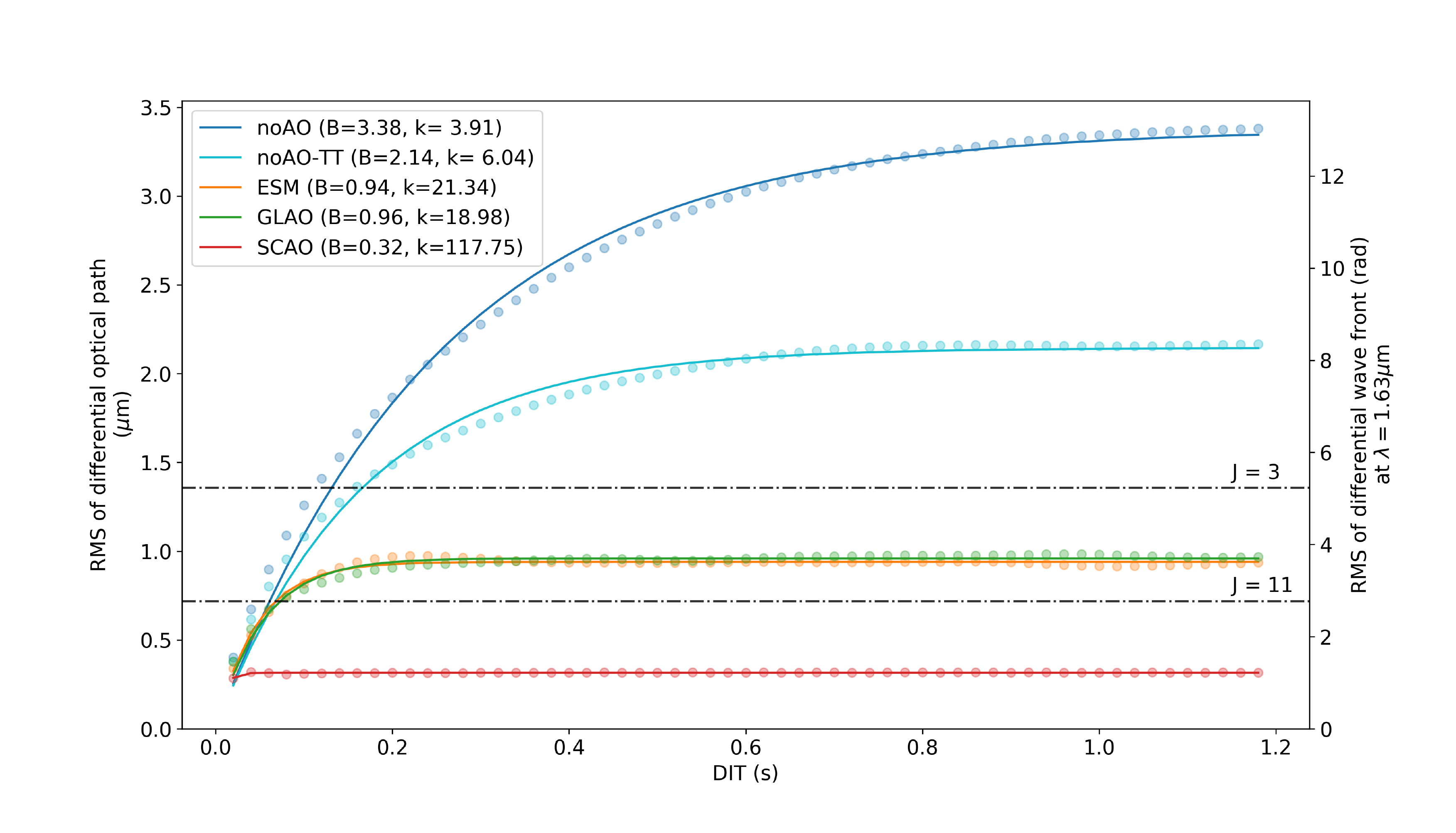}
	\caption{Mean aperture rms of the wavefront differentials $\Delta\phi(\Delta t) \equiv \phi(t) - \phi(t + \Delta t)$ as a function of the time interval $\Delta t$. The solid lines indicate the bounded-growth fit to the respective curve (see Eq.~\ref{eq:bounded_growth_fit}) and the dash-dotted black lines indicate $\sqrt{2} \times$ the expected values for the atmospheric wavefront error after correcting for increasing radial Zernike orders, based on the results of \cite{Noll76}. The parameters of the bound-growth fit are given in the legend, where $B$ in $\mu$m and $k$ in s$^{-1}$.}
	\label{fig:divergencetrends}
\end{figure*}

As mentioned above, the boundary $B$ represents the mean difference between two randomly selected wavefront planes and serves as a measure of the residual power in the atmospheric turbulence spectrum, the Kolmogorov or van-Karman spectrum. The application of an AO control reduces the power of the aberration spectrum and, thus, these boundary limits are expected to decrease with increasing maximum controlled Zernike order. As we compute the difference of two randomly selected wavefronts, the variance expectation value from \cite{Noll76}, $\Delta_J$, has to be doubled and, hence, the rms expectation value is multiplied by a factor of $\sqrt{2}$, see the dash-dotted line in Fig.~\ref{fig:divergencetrends}.

The corresponding results from evaluating the wavefront data are presented in Fig.~\ref{fig:divergencetrends} for each simulation setup.
As expected, we see that the boundary values decrease with increasing order of correction. The uncontrolled wavefront (noAO, blue curve) reaches the boundary plateau at DITs after $\sim1.2$\,s, where the controlled wavefronts reach the plateau after shorter time intervals, which is due to the fact that the slowly varying low-order aberrations are filtered out by the AO system and the fast varying higher-order aberrations are uncorrelated after shorter time scales. The growth constants, $k$, translate into characteristic time scales of 255.8, 165.6, 46.9, 52.7, and 8.5\,ms, respectively.

\begin{figure*}
	\centering
	\includegraphics[width=.95\textwidth]{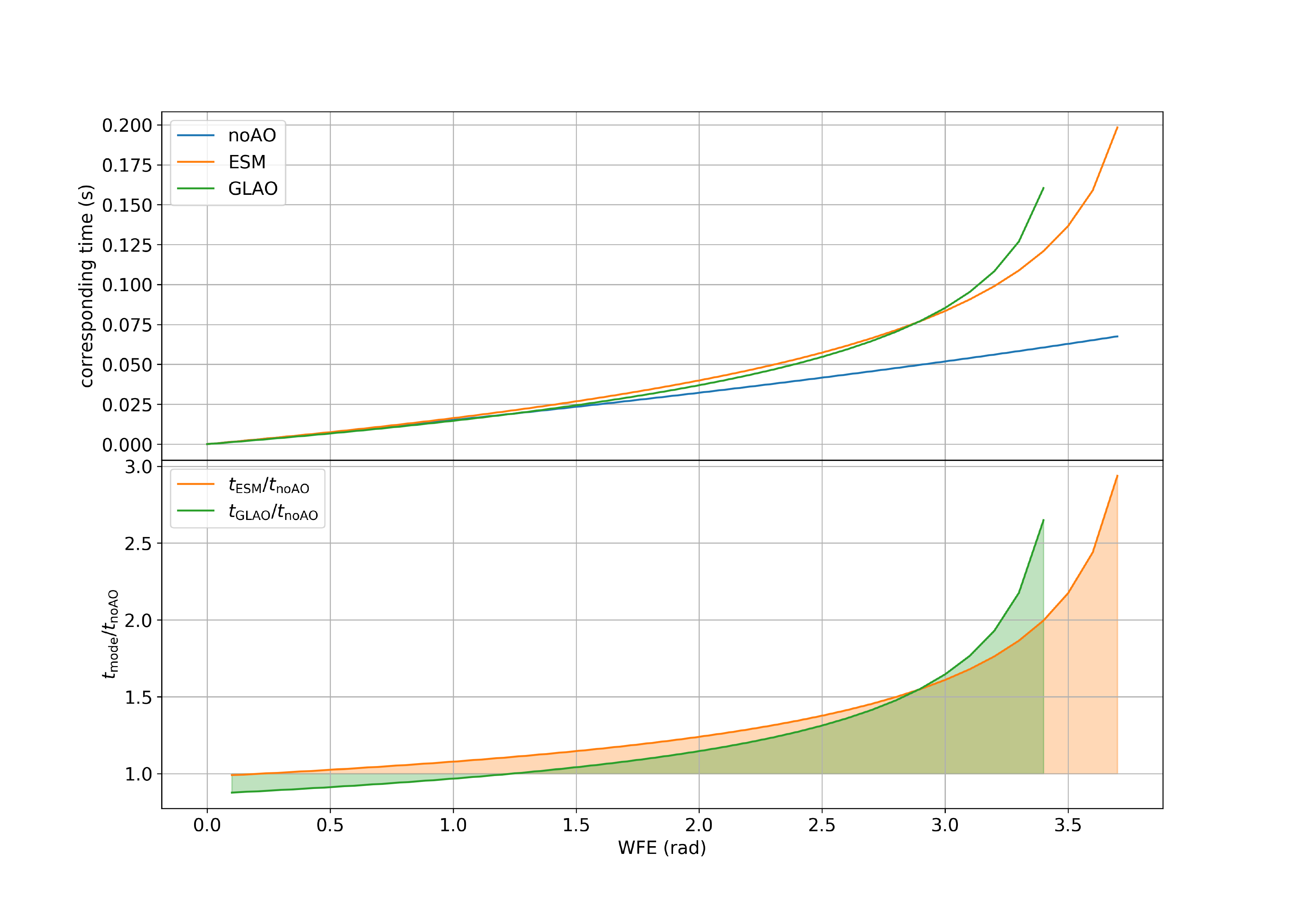}
	\caption{(\emph{Top}) Average time required to reach a certain wavefront decorrelation at $\lambda = 1.63\mu$m for three types of wavefront control. The required time for the GLAO control grows to infinity before 4\,rad since the control limits the maximum decorrelation to $\sim3.5$ rad and is therefore not plotted for higher values. (\emph{Bottom}) Average gain in achievable integration time of AO control versus the uncontrolled wavefront (noAO). The gain diverges towards infinity for larger wavefront error budgets.}
	\label{fig:DIT_differences}
\end{figure*}

Furthermore, we see that the controlled wavefronts require longer integration times to decorrelate to a given mean RMS value of the differential wavefront. This will allow for increasing the integration times of the short exposures for the imaging process, and thus be beneficial to achieve higher SNR or allow to read larger detector read-out areas\footnote{We note that preceding applications of holographic imaging were based on data obtained from windowed detectors to increase the achievable read-out speed.} (thus larger FoV) in a given amount of time. To quantify this behavior, we compare the time required to reach a given wavefront decorrelation for the noAO, ESM and GLAO wavefronts, see upper panel in Fig.~\ref{fig:DIT_differences}, by basically flipping the $x$ and $y$-axes of Fig.~\ref{fig:divergencetrends}. The bottom panel compares these times of the controlled wavefronts (ESM \& GLAO) to the noAO case. The advantage of the control is obvious at wavefront error levels $\gtrsim 3.0$\,rad for the GLAO and ESM cases, respectively, where the graphs diverge towards positive infinity since the values for the controlled cases are limited $\lesssim 4.0$\,rad.

However, at such long integration times or such large wavefront errors the contrast in the PSF is almost gone, since a wavefront RMS of 1, 2, and 3\,rad corresponds to a mean fringe contrast/ Strehl loss down to approximatively 60\%, 14\% and 1\%, respectively. \cite{Schoedel13} found that using short exposures, being integrated significantly longer than the atmospheric coherence time $\tau_0$, still allows for reaching the diffraction limit but at the cost of a lower Strehl ratio, as information is lost due to the loss of contrast. The curves now suggest that we will achieve the same Strehl even though the integration times for the short exposures are extended by a factor of 2, if we accept an RMS value of $\sim3.2$\,rad and $\sim3.4$\,rad in the two cases, respectively.

\label{sec:longer_integration_times}
Apart from that, the longer integration time will increase the SNR of the PSF estimate during the holographic reconstruction and therefore will probably increase the recovered Strehl. Especially the high spatial frequencies, relevant to achieve a high Strehl, are read-noise limited for short exposures, hence a doubled integration time will deliver a doubled SNR, or allow to observe stars 0.75\,mag fainter at the same SNR. We will analyze this in the following section.

\section{Signal-to-noise ratio in the simulated data}
\label{sec:signaltonoise}
The technique of holographic imaging is based on the result by \cite{Primot90}, that the best least-squares estimate of the Fourier transform of an object $O$ is given by Eq.~\ref{eq:reconstructionformula}, where the $m$-th short exposure image $I_m = O \ast P_m$ is the result of a convolution (as denoted by the in-line asterisk $\ast$) of the object with the instantaneous PSF $P_m$.
\begin{eqnarray}
	\mathcal{F}O &= \frac{\left\langle \mathcal{F}I_m \cdot \mathcal{F}P^*_m \right\rangle}{\left\langle \mathcal{F}P_m \cdot \mathcal{F}P^*_m \right\rangle} \label{eq:reconstructionformula}\\
		&= \frac{\left\langle \mathcal{F}I_m \cdot \mathrm{OTF}^*_m \right\rangle}{\left\langle \mathrm{MTF}^2_m \right\rangle} \label{eq:reconstructionformula_subst}
\end{eqnarray}
In these expressions, the $^\ast$ denotes the complex conjugate and the averages, $\left\langle \cdot \right\rangle$, are taken over all $M$ short exposure images. In the second expression, we substitute the Fourier transform of the PSF by the equivalent \emph{optical transfer function}, OTF$= \mathcal{F}P$, whose absolute value is the \emph{modulation transfer function}, MTF = $|$OTF$|$. For the following analysis, it is very useful that squaring this function, $\mathrm{MTF}^2$, yields the power spectrum of the PSF. From Eq.~\ref{eq:reconstructionformula_subst}, we directly see that the quality of the holographic reconstruction depends on the measurement of the PSF power spectrum. Therefore we analyze the signal-to-noise ratio (SNR) of MTF$^2$ in the synthetic observations in the following by varying a set of parameters. We emphasize here that the following quantitative results depend to a significant extent on the actual atmosphere, its vertical $C_n^2$ and wind speed profile.

\subsection{Integration time}
\begin{figure*}
	\centering
	\includegraphics[width=.95\textwidth]{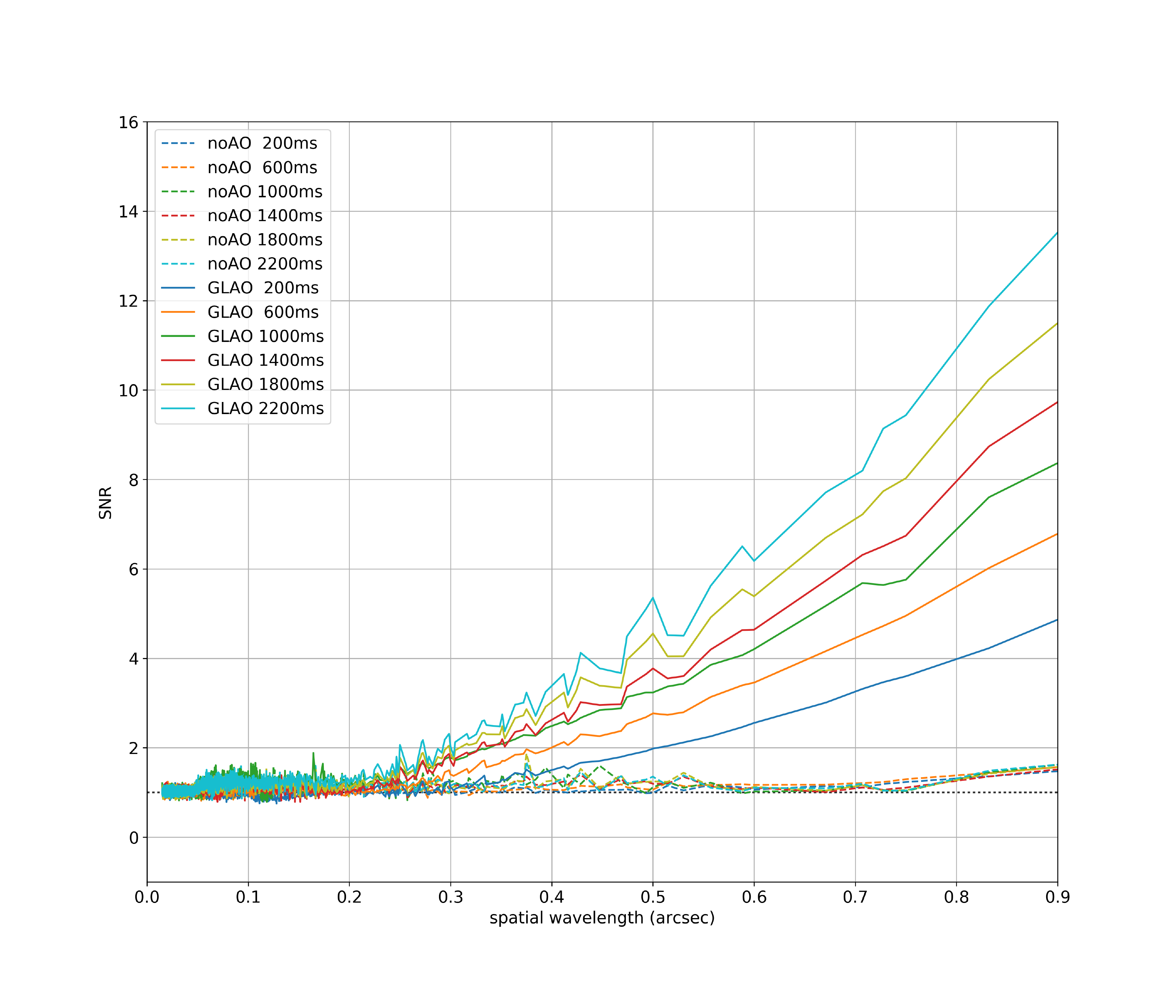}
	\caption{SNR of the power spectra for varying integration times and wavefront controls as a function of spatial wavelength in the aperture. The solid lines are for GLAO control, and the dashed lines are for the noAO case. The colors are shared for the same integration time, see the legend.}
	\label{fig:power_spectra_integration_time_comparison}
\end{figure*}

Longer integration times of the short exposures will increase the SNR of the PSF estimate, as mentioned in Sect.~\ref{sec:wavefrontdecorrelation}. This is supposed to increase the Strehl ratio in the reconstructed image and therefore we compare the SNR in the power spectra for varying integration times in Fig.~\ref{fig:power_spectra_integration_time_comparison}. All curves are measured within an aperture with a radius of 1.5\,arcsec, around a $H=12$\,mag star.

We find that the SNR increases as a function of integration time as expected, in particular at the long spatial wavelengths due to their longer coherence times. Furthermore, the measurements in the GLAO data tend to yield a higher SNR, where this effect is most prominent for spatial wavelengths larger than 0.25\,arcsec.
This suggests that, from a SNR point of view, exposures should be taken with as long as possible integration times, to beat down the noise contributions, and that the application of the GLAO correction is increasing this effect significantly for the longer spatial wavelengths.

\subsection{Brightness of reference stars}
\begin{figure*}
	\centering
	\includegraphics[width=.9\textwidth]{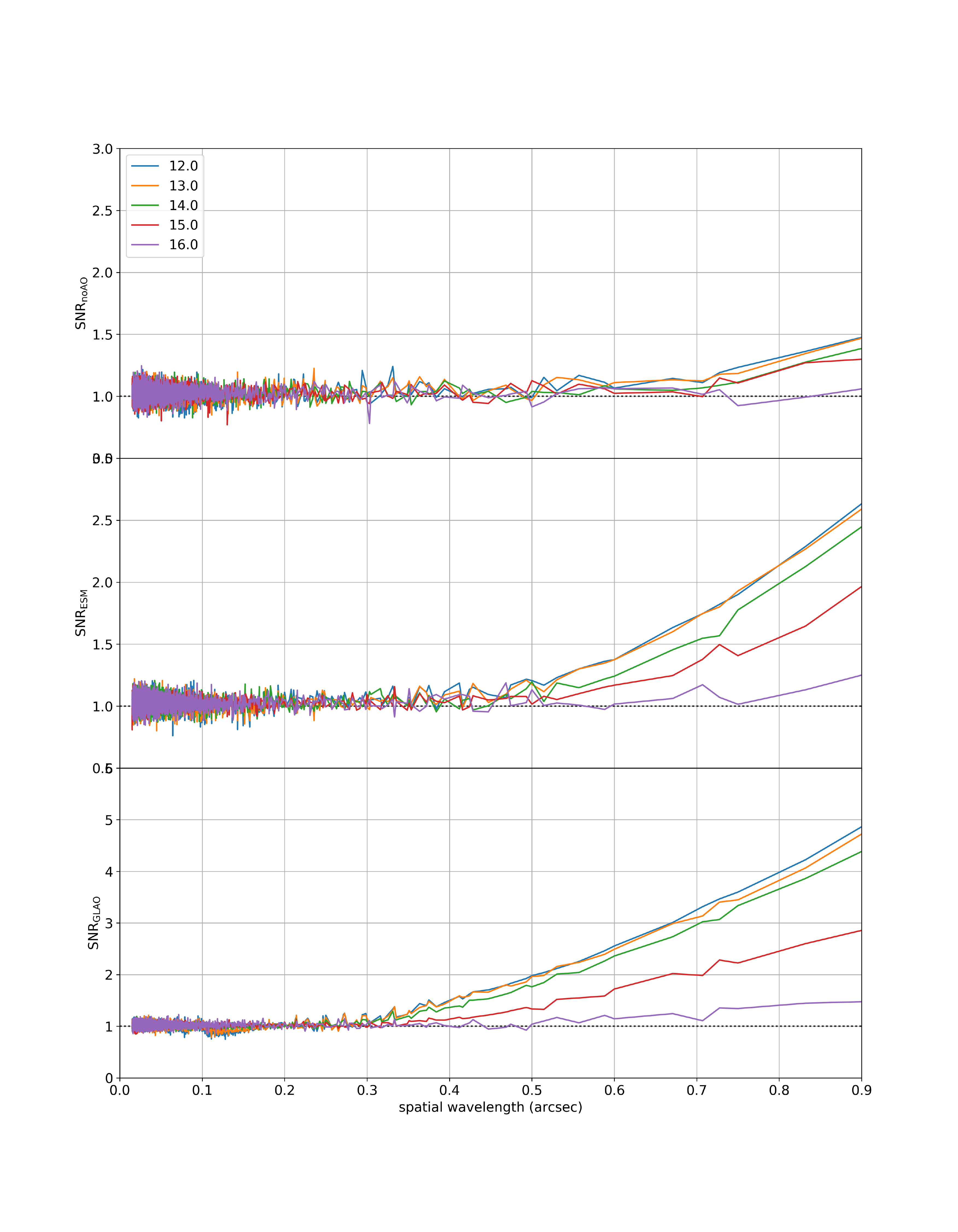}
	\caption{SNR of the power spectra for stars with different $H$-band magnitudes as a function of spatial wavelength in the aperture. The plots show the same family of curves for three types of wavefront control (top: noAO, mid: ESM, bottom: GLAO) for the aperture radius of 1.5\,arcsec and for DIT\,=\,200\,ms.}
	\label{fig:power_spectra_magnitudes}
\end{figure*}

\begin{figure*}
	\centering
	\includegraphics[width=.95\textwidth]{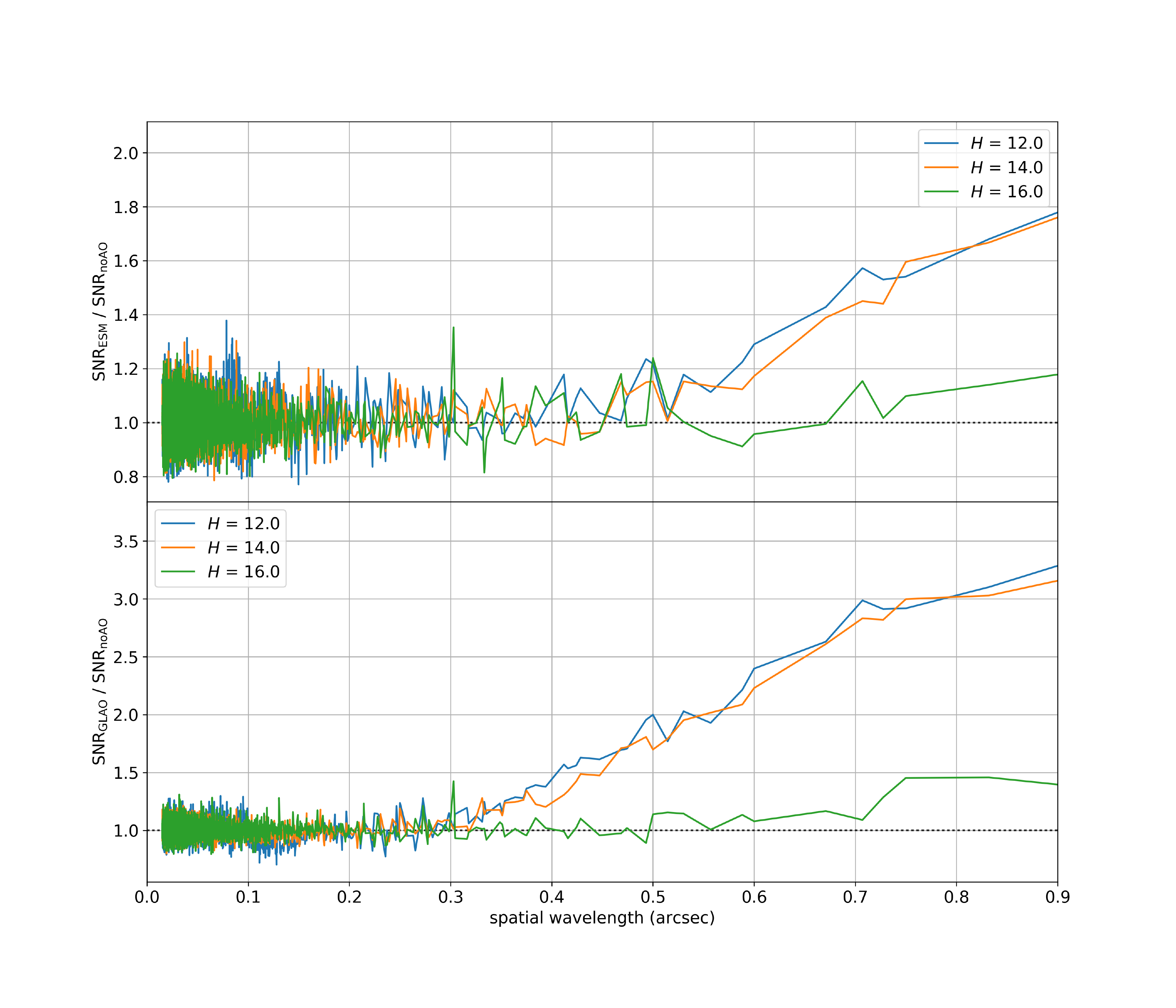}
	\caption{Gain in the SNR of the power spectra for a ESM (\emph{top}) and GLAO (\emph{bottom}) wavefront control versus no control (noAO), as a function of spatial wavelength in the aperture and for three stellar magnitudes. The dotted black line indicates a ratio of order unity.}
	\label{fig:power_spectra_mode_gain}
\end{figure*}

A second important parameter contributing to the achievable reconstruction quality is the magnitude of the reference star(s). \cite{Schoedel13} found that a group of faint reference stars ($K_s = 13 \pm 0.5$) may achieve a similar or even better result than using a single bright star ($K_s = 12$), where using multiple reference stars in a crowded field reduces the systematic sources of uncertainty, i.e. this (i) increases the SNR of the PSF estimate per frame and (ii) also takes into account the variation of the PSF across the FoV. But in this paper we confine ourselves to compare only the power spectra for different stellar magnitudes, see Fig.~\ref{fig:power_spectra_magnitudes}, and \ref{fig:power_spectra_mode_gain}, since the variability of the PSF across the FoV is significantly reduced by the (GL)AO correction, anyways.

Fig.~\ref{fig:power_spectra_magnitudes} surprisingly suggests that choosing reference stars brighter than $H \approx 14$\,mag does not result in higher SNR per spatial wavelength mode. However, there is an obvious increase of the SNR of up to a factor of 1.6 or 3.0 for the spatial wavelengths larger than 0.5\,arcsec, when using the wavefront control of the ESM and GLAO systems, respectively. This is more prominent in the subset of curves in Fig.~\ref{fig:power_spectra_mode_gain}.
The curves in Fig.~\ref{fig:power_spectra_magnitudes} suggest that a $H \approx 15$\,mag allows for a higher SNR of 2 for the long spatial wavelength regime, when the GLAO correction is applied, which is not even reached for $H \approx 12$\,mag stars when not applying the AO correction. This strongly suggests, that the application of the GLAO correction allows for still getting significant SNR when using much fainter holography-reference-stars, of about $\Delta H = 3$\,mag, for integration times of 200\,ms. This furthermore enables the usage of more (fainter) reference stars, resulting in a furthermore decreased noise level. The ESM allows for an intermediate increase of SNR per spatial wavelength mode, where the homogeneity of the PSFs across the FoV needs to be studied in more detail, leaving the GLAO correction as the favourite mode.

\subsection{Aperture radius}
\begin{figure*}
	\centering
	\includegraphics[width=.85\textwidth]{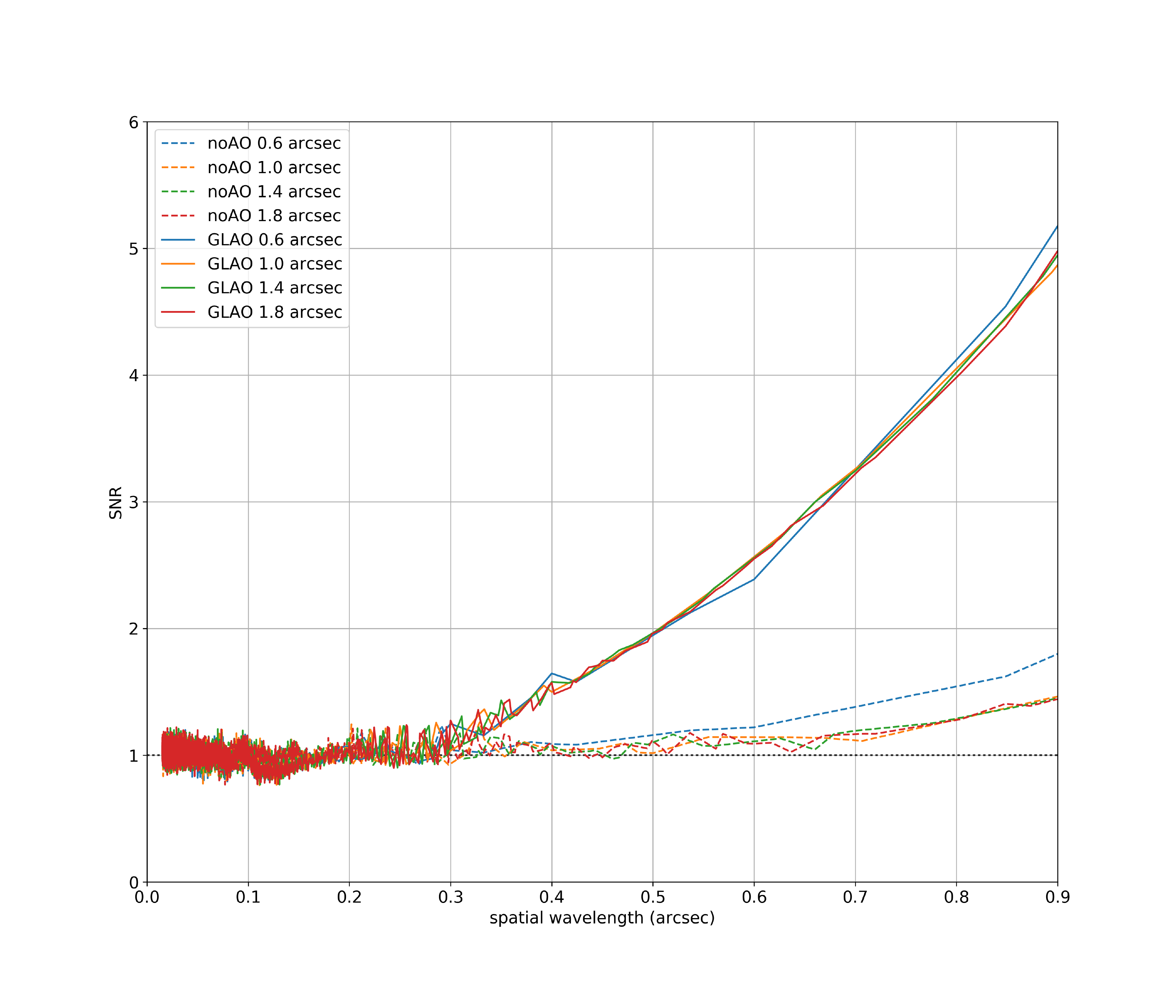}
	\caption{SNR of the power spectra for the same star when varying the reference aperture radii as a function of spatial wavelength in the aperture. The plot shows the same family of curves for two types of wavefront control and for the DIT of 200\,ms.}
	\label{fig:power_spectra_aperture_radii}
\end{figure*}

Finally we tested how the choice of the aperture radius affects the measurement. Therefore, we applied our analysis to the same data set and varied the radius of the aperture over which we measured the SNR, while keeping the DIT fixed. We find in the resulting Fig.~\ref{fig:power_spectra_aperture_radii} that choosing a larger aperture radius does not affect the SNR per spatial wavelength mode as the curves for the same data set are overlapping nicely.

\section{Holographic image reconstruction of the synthetic observations}
\label{sec:holographicimagereconstruction}
\begin{figure*}
    \centering
    \includegraphics[width=.85\textwidth]{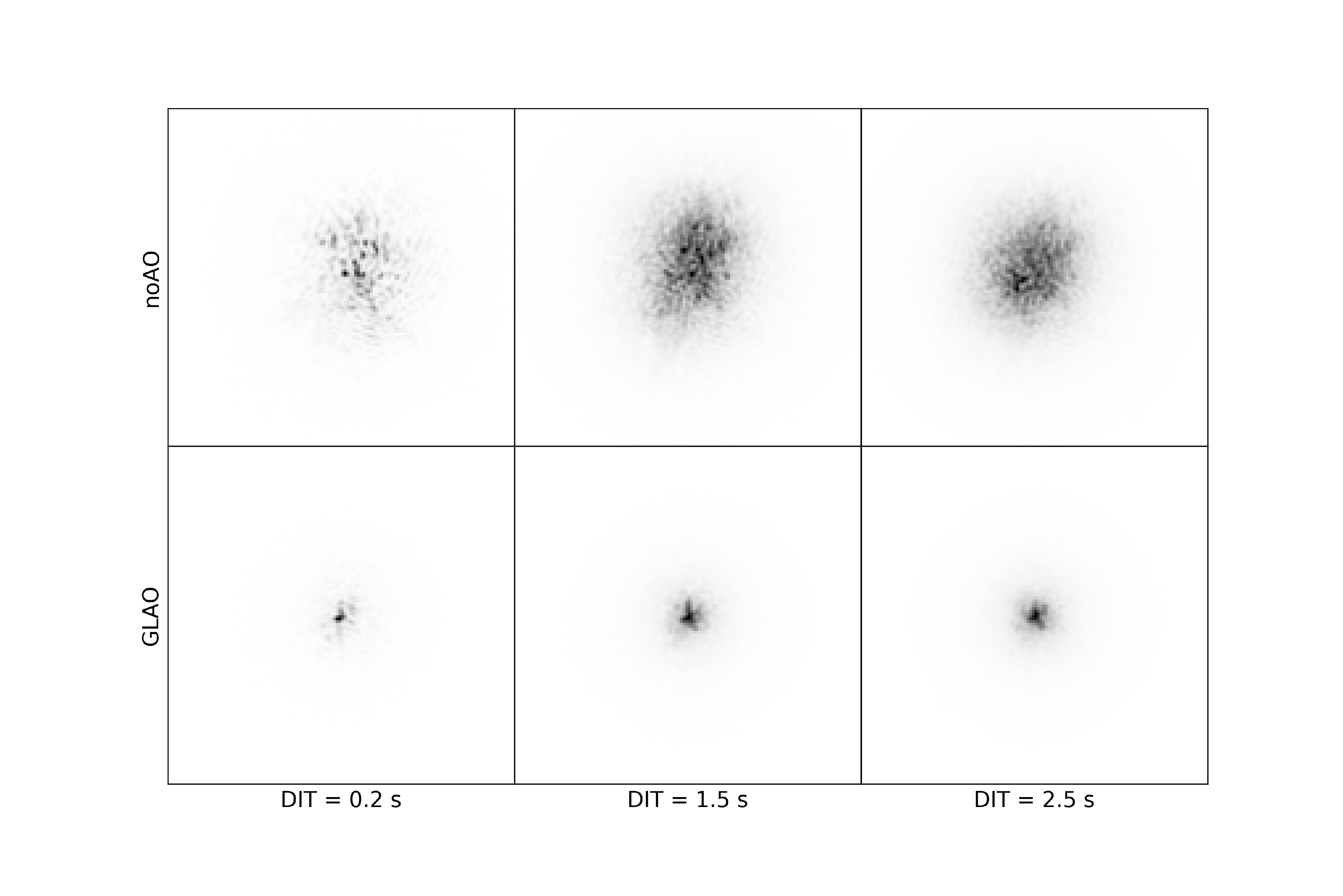}
	\caption{Example PSFs from the noAO (upper panels) and GLAO simulations (lower panels). The PSF data are integrated to a total DIT of 0.2, 1.5, and 2.5\,s, respectively, as used in the synthetic observations, IDs~2--6. All PSFs are normalized to an integrated flux of order unity.}
	\label{fig:example_psfs}
\end{figure*}

We test the predictions from above by applying the method of holographic imaging on the synthetic observations from Sect.~\ref{sec:simulations}. Therefore, we use the reduction pipeline from \cite{Schoedel13} and apply it to the data sets from Table~\ref{tab:observations}. These sets of short exposures contain 800 frames each and have a total integration times of 160, 1200, and 2000\,s. Examples of the flux-normalized PSFs are given in Fig.~\ref{fig:example_psfs}. In these images, there are two prominent features, first the spread of the speckle cloud in the noAO simulations, which is not apparent in the GLAO data, and second smoothing of the individual speckles towards longer DITs, which presumably limits the recoverable Strehl and disables the disentanglement of close sources.
The same four bright stars are chosen as reference stars to obtain a comparable reconstruction, only the noAO data set with DIT\,=\,200\,ms was restricted to the two brightest stars because the fainter stars did not have sufficient SNR to improve the PSF estimation with such short exposure times.

\begin{table*}
	\centering
	\caption{Observations used in this work.}
	\label{tab:observations}
	\begin{tabular}{rllllll}
	\hline \hline
	ID	& Target			& Instrument		& Band  & Seeing (arcsec)$^a$& DIT (s)	& $N_\mathrm{Frames}$ 	\\
	\hline
	1	& Galactic Center	& VLT/NaCo			&       & 0.36 (0.28) 		& 0.15 		& 500				\\
	2	& \multicolumn{2}{c}{synthetic noAO}	& $H$   & 1.07 (0.84)		& 0.2		& 800 				\\
    3	& \multicolumn{2}{c}{ }					& $H$   & 					& 1.5		& 800 				\\
	4	& \multicolumn{2}{c}{synthetic GLAO}	& $H$   & 0.44 (0.35)$^b$	& 0.2		& 800				\\
	5	& \multicolumn{2}{c}{ }					& $H$   & 					& 1.5		& 800				\\
   	6	& \multicolumn{2}{c}{ }					& $H$   & 					& 2.5		& 800				\\
   	7   & $\gamma$~Vel      & VLT/HAWK-I        & $K_s$ & --                & 2.0       & 500               \\
   	8   & $\gamma$~Vel      & VLT/HAWK-I        & $Y$   & --                & 2.0       & 250               \\
	\hline
	\end{tabular}
	\\\textbf{Notes:} $^a$ The seeing is given for optical ($H$-band) wavelengths. $^b$ The seeing estimate for the GLAO observations is estimated after the AO correction, the atmospheric input was the same as for the noAO observations.
\end{table*}

\begin{figure*}
	\centering
	\includegraphics[width=.95\textwidth]{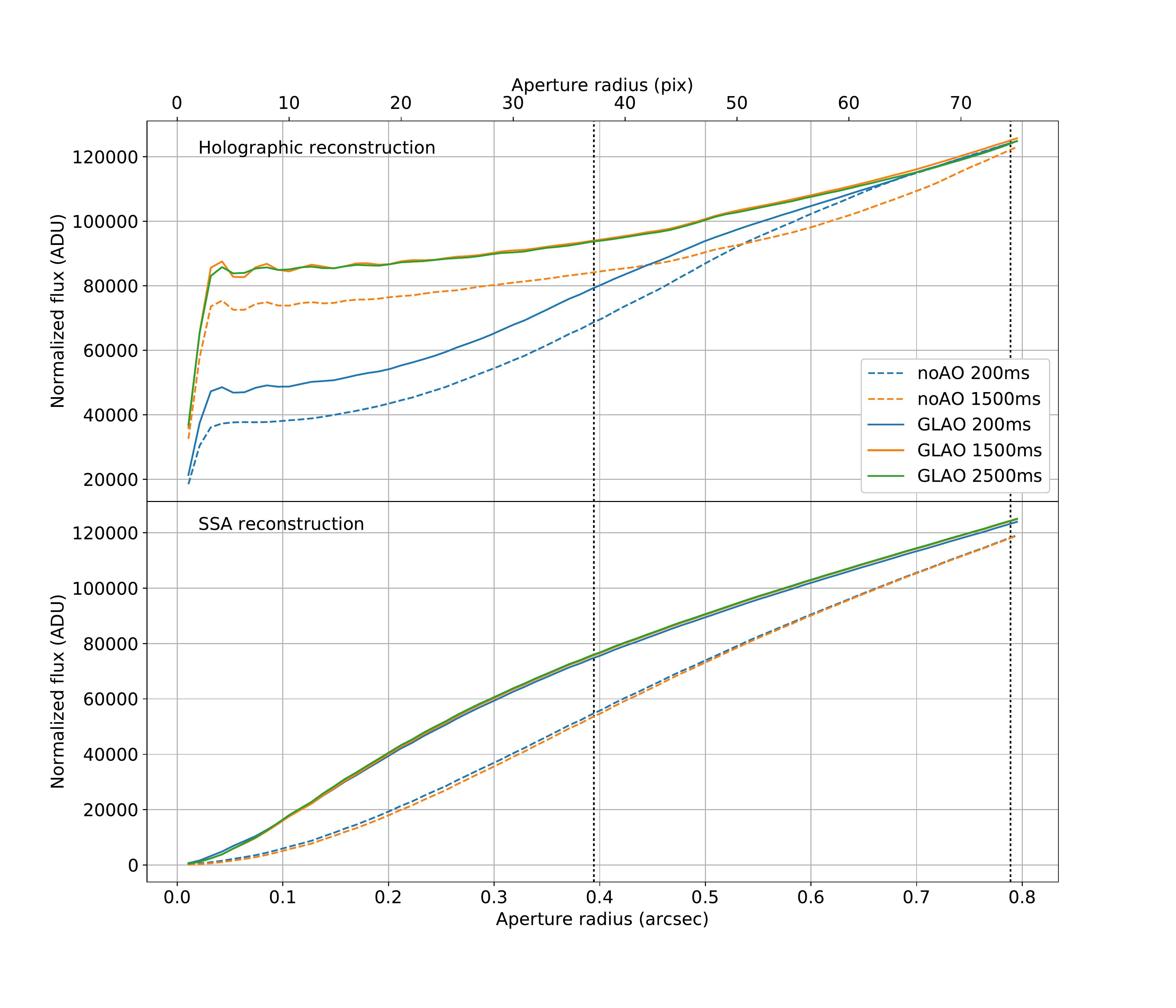}
    \caption{Encircled energy in the reconstruction of the data sets around a $H=12.4$\,mag star, in the holographic reconstruction (\emph{top}) and in the simple-shift-and-add reconstruction (\emph{bottom}). The flux is normalized to the total integration time of 160\,s. The dotted vertical lines indicate one and two times the expected HWHM, corresponding to $R_\mathrm{aperture} = 0.4$ and~0.8\,arcsec, respectively.}
    \label{fig:encircled_energy}
\end{figure*}

We compare the curves of encircled energy for the brightest star in the synthetic cluster ($H=12.4$\,mag) in Fig.~\ref{fig:encircled_energy}. In this plot, the flux is normalized to the identical total integration time of 160\,s, corresponding to the 200\,ms per frame data sets.
From these curves we directly see, that the application of a GLAO correction shifts more flux to the central peak, as seen in the respectively steeper rise of the curves. This suggests, that this mode is favored over observations without AO correction. Towards larger aperture radii, the curves converge towards the radial distance where all stellar flux is encircled and the additional flux is from the sky background only. From this point on, the curves overlap, as expected. We identify the same behavior towards fainter stars in the field, regardless of whether they are reference stars or not. The curves for fainter stars converge at smaller radii what is expected as they do not contribute significant amounts of flux towards the larger radial distances, or their additional flux at such large radial distances is comparable to the sky background.

Furthermore, there is a prominent difference between the curves from different integration times. The longer integration times apparently result in a better reconstruction, as suggested in Sect.~\ref{sec:signaltonoise}, especially in Fig.~\ref{fig:power_spectra_integration_time_comparison}. However, this is no longer true for the longest integration times as the time-normalized GLAO curves for DIT\,=\,1500\,ms and~2500\,ms overlap quite well, what suggests that the gain in the SNR is equalized by some other effect, but this also confirms the finding from Sect.~\ref{sec:wavefrontdecorrelation} that the reconstruction is expected not to suffer from longer integration times beyond the decorrelation time scale.

Comparing this to the results from the simple shift-and-add (SSA) algorithm \citep{Bates80} in the bottom panel, we directly see that the energy is spread over a larger area than in the holographic reconstructions, but still the GLAO-assisted observations provide a better reconstruction, as more energy is focused towards the center. The quality of the SSA reconstruction appears to be independent on the DIT. We note that the SSA reconstruction is based on the first 100 frames of each data set, but normalized to the same total exposure time.

\section{Comparing speckle holography with SSA on VLT/HAWK-I data with GLAO correction}
\label{sec:holography__vs_ssa}
Speckle holography can also be used to improve the sharpness of images without the necessity to go to the diffraction limit, for example if the sampling of the detector limits the angular resolution, as is the case with VLT/HAWK-I, with a pixel scale of 0.106\,arcsec. The GALACTICNUCLEUS survey uses this technique to obtain 0.2\,arcsec FWHM images at $JHK_s$ with HAWK-I and short exposure times \citep{NoguerasLara18}. In this section, we test speckle holography on HAWK-I $K_s$ and $Y$ data (IDs 7 and 8 in Table~\ref{tab:observations}), obtained with the ground-layer correction of the VLT/AOF 4~Na-LGS~AO subsystem GRAAL, and compare the result to the standard SSA image reduction. Target of these observations was the nearby (350\,pc) and young ($\sim$7\,Myr old) star cluster $\gamma$-Velorum.

The $K_s$ ($Y$) data consist of 20 (10) cubes of about 25 exposures of 2\,s DIT each. We only analyzed the data from a single one of HAWK-I's four detectors. We extracted the PSF from each individual frame by superposing the images of seven bright, isolated stars distributed across the field. We found the PSF to be homogeneous across the entire $2\times2$\,arcmin FoV. This facilitated the application of the holography algorithm considerably and demonstrates the high quality of the GRAAL + AOF system.

\begin{figure}
	\centering
	\includegraphics[width=.52\textwidth]{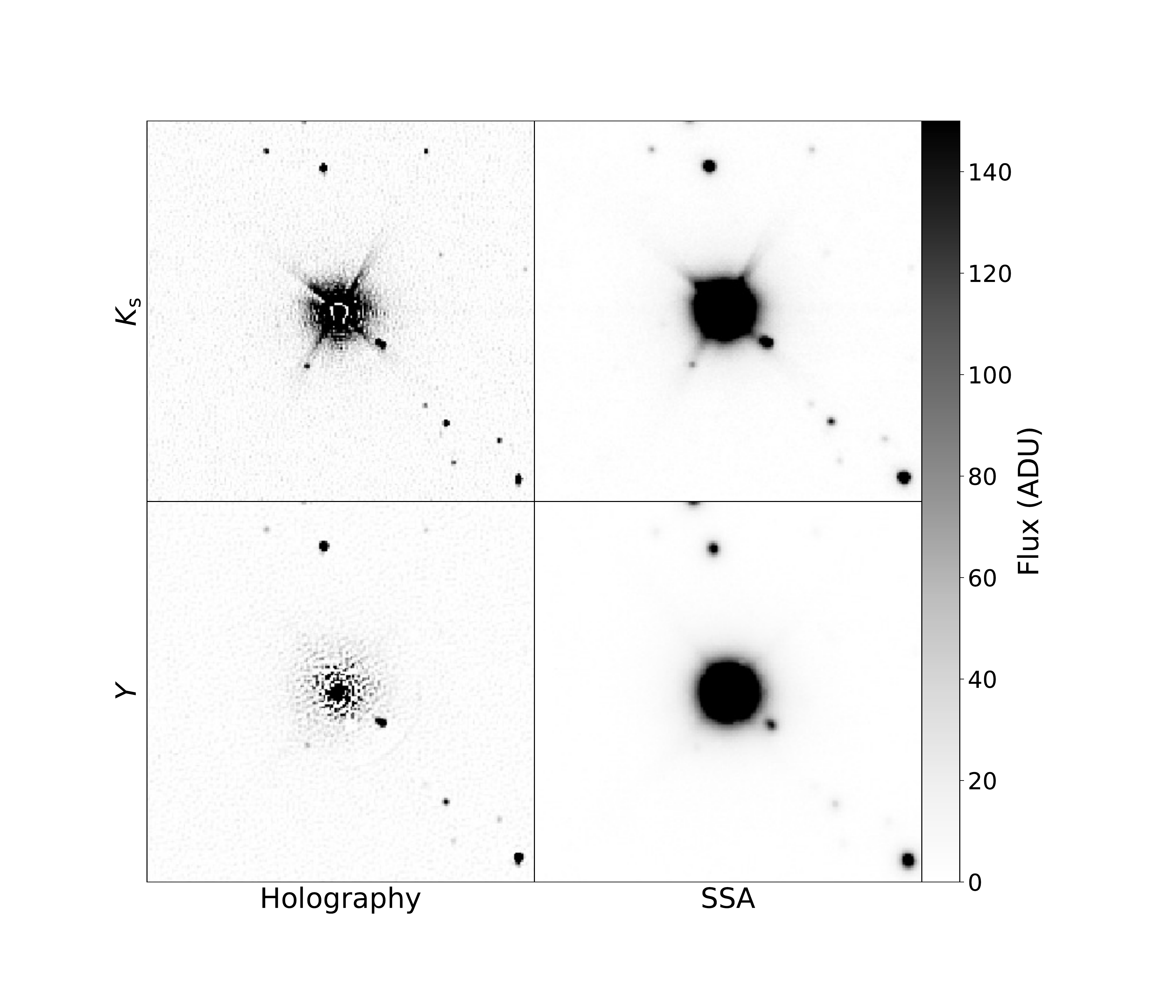}
	\caption{The panels show a 21.2\,arcsec by 21.2\,arcsec subfield of the VLT/HAWK-I+ GRAAL data, centered on the brightest star, in the holographic and SSA reconstruction for the two filter bands, as indicated.}
	\label{fig:reconstruction_psfs}
\end{figure}

We used a Gaussian of 0.25\,arcsec FWHM to create the final holographic images (cf. the extracts in Fig.~\ref{fig:reconstruction_psfs}). To ensure robust source detection and accurate assessment of photometric and astrometric uncertainties we applied a bootstrap procedure to both the SSA and holography data reduction. From the each of the $K_s$ and $Y$ data cubes we created one individual image and then randomly selected 20 (10) of those images (resampling with replacement, so any given image can be a repeated one or several times) and created deep mean images and corresponding noise maps. We thus created 21 resampled deep images. Those were then analyzed with the \textsc{StarFinder} software \citep{Diolaiti00}, with a correlation threshold of 0.7, two iterations with $3\sigma$ detection limits, and deblending blurred stars. The 21 star lists were then combined. Stars detected within 2 pixels of each other were considered to be the same star. Finally, to avoid spurious detections, we required a star to be detected in 90\% of the resampled images. The fluxes and positions of the stars, as well as their uncertainties, were taken from the mean and standard deviation of the individual measurements.

The holographic images are significantly sharper than the SSA images. However, the quality of the AO correction of the data is so good, that all very close stars were disentangled by \textsc{StarFinder} even in the SSA images. More stars were detected in the SSA images (10\% and 20\% for $K_s$ and $Y$, respectively). The missing stars in the holography images are all at the faint end of the luminosity function. This is probably due to the presence of correlated  noise in the holographically reduced images, which leads to a graininess of the background that has a scale on the order of the FWHM of the stars and can thus hinder the detection of faint stars (cf. Fig.~\ref{fig:reconstruction_psfs}). Holography requires a large number of frames to beat down the noise in the denominator (see Eq.~\ref{eq:reconstructionformula}) and the number of frames used here is relatively small, in particular at $Y$.

\begin{figure*}
	\centering
	\includegraphics[width=.45\textwidth]{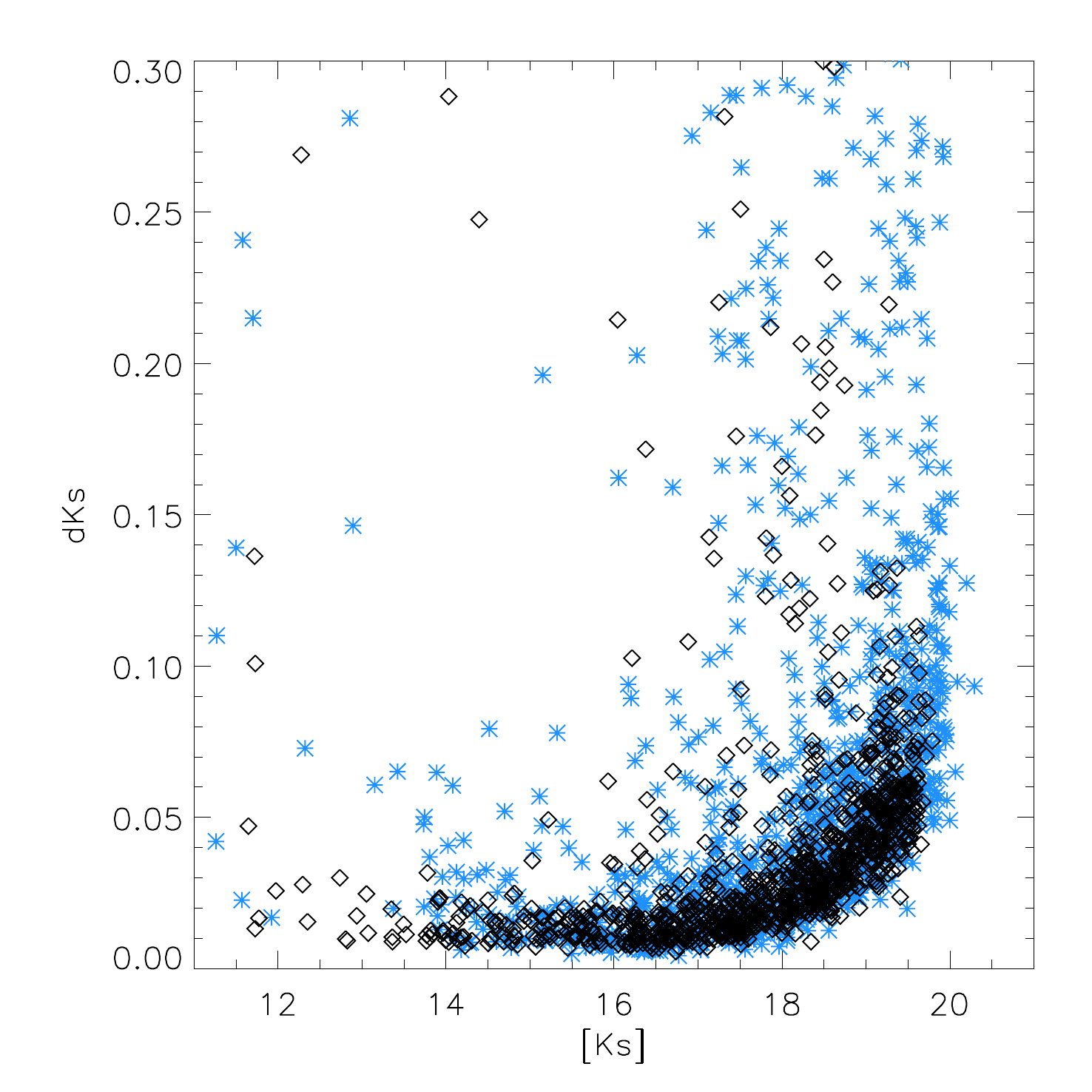}
	\includegraphics[width=.45\textwidth]{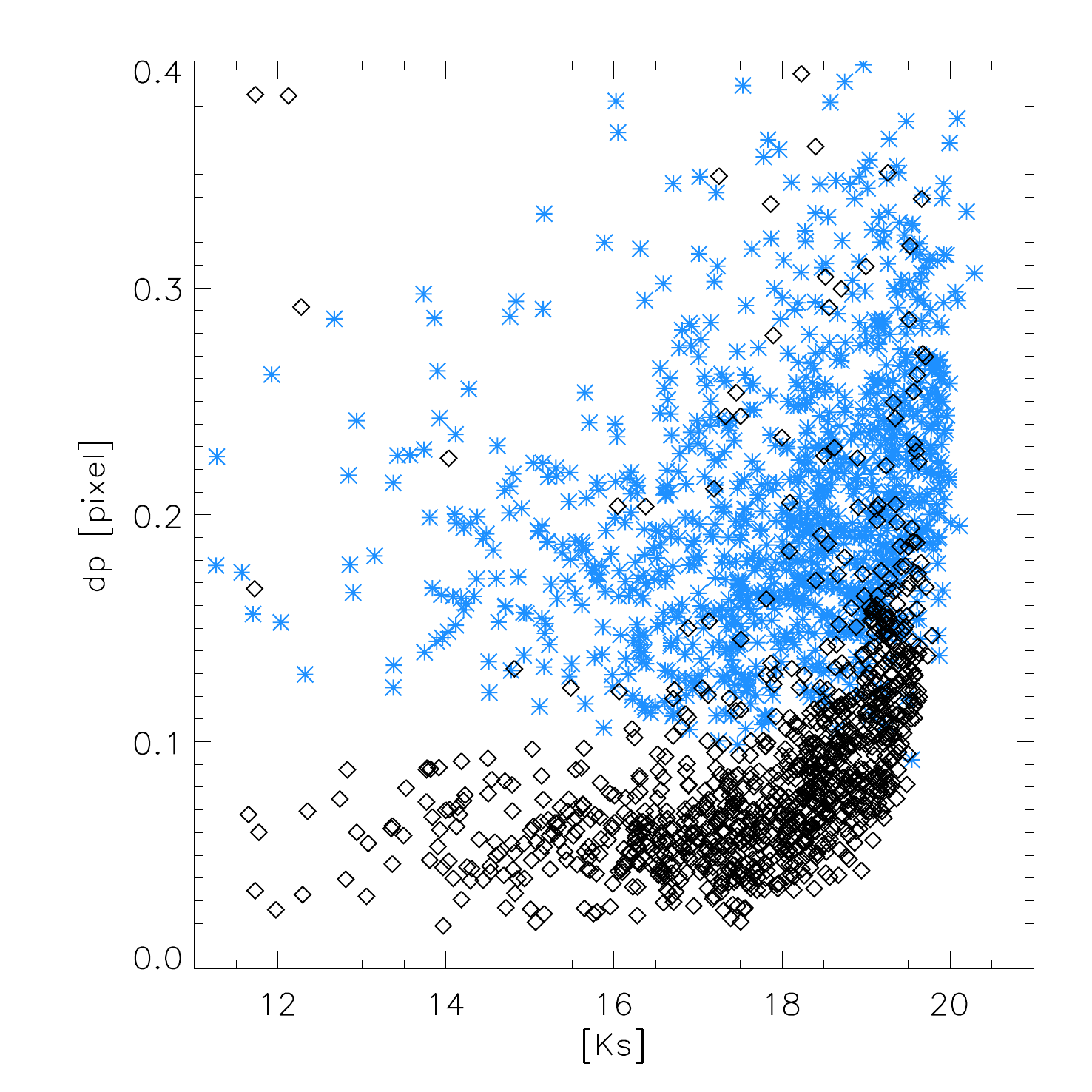}
	\caption{Photometric (\emph{left}) and astrometric (\emph{right}) uncertainties over $K_s$ magnitude as identified by \textsc{StarFinder} in the VLT/HAWK-I+GRAAL data (IDs 7 and 8 in Table~\ref{tab:observations}). Blue stars denote the uncertainties in the SSA reduction, and black diamonds the corresponding holographic reduction.}
	\label{fig:hawki+graal}
\end{figure*}

On the other hand, both the photometric and astrometric uncertainties of the detected stars are significantly smaller in the holography images, see Fig.~\ref{fig:hawki+graal}. We believe that this is related to the internal algorithms of the \textsc{StarFinder} software. \textsc{StarFinder} only uses the cores of the stars to fit their position and flux. The SNR of the cores of the bright stars is much higher in case of the holography images.
\begin{figure*}
	\centering
	\includegraphics[width=.95\textwidth]{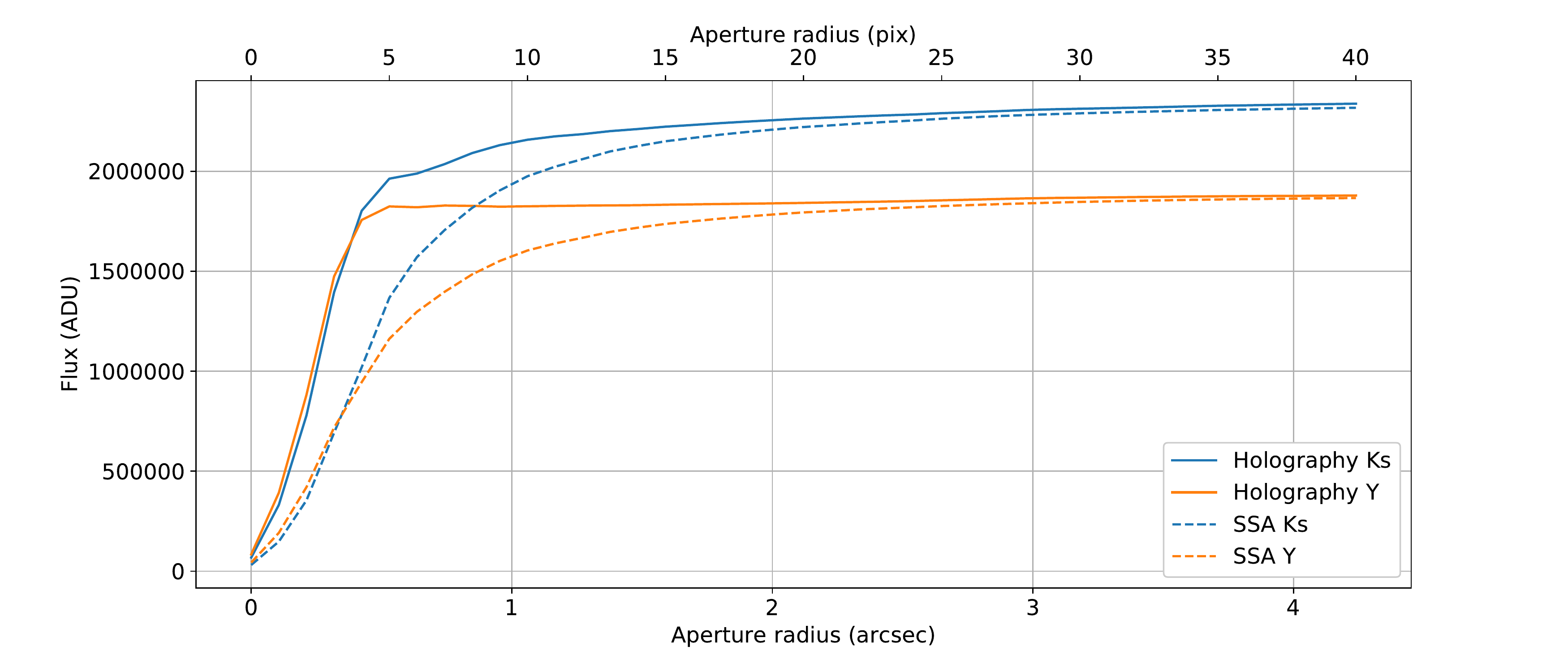}
	\caption{Encircled energy in the reconstruction of the VLT/HAWK-I+ GRAAL data around the brightest star in the field. Solid lines indicate the holographic reconstruction and dashed lines the simple-shift-and-add reconstruction. Blue and orange indicate the $K_\mathrm{s}$ and $Y$ band, respectively.}
	\label{fig:encircled_energy_obs}
\end{figure*}
In Fig.~\ref{fig:encircled_energy_obs}, we show the graphs of encircled energy for the reconstructions presented in Fig.~\ref{fig:reconstruction_psfs}. It is clearly visible that the holography method focuses the energy to a region a factor of 2 smaller than in the SSA reconstruction. Moreover, in contrast to the SSA technique, the holographic approach will likely yield a much sharper reconstruction for data taken with a detector with a much smaller pixel scale, sampling the diffraction limit of the telescope (cf. Fig.~\ref{fig:encircled_energy}).

In summary, the astrometric uncertainties in the holographic reconstruction are about a factor of 2 lower than in the SSA reconstruction (cf. Fig.~\ref{fig:hawki+graal}), even though the observational data have a limited spatial resolution (due to the VLT/HAWK-I pixel scale of 0.106\,arcsec). Besides the higher detection limit of $\sim0.5$\,mag (due to the low number of short-exposure frames), this analysis clearly prefers the holographic to the SSA reconstruction technique for obtaining diffraction (or pixel scale)-limited imaging. However, we need speckle observations from an instrument with a pixel scale sampling the diffraction limit of the telescope, to fully characterize the advantage of the holography technique.

\section{Summary and conclusions}
\label{sec:summary}
In this paper, we analyzed the potential of applying the holographic image reconstruction algorithm to AO-assisted short exposure observations from 8\,m class telescopes. We simulated series of point spread functions for natural seeing (noAO), ground layer AO (GLAO), single-conjugate AO (SCAO) and the low-order enhanced seeing mode (ESM), available for LBT/LUCI observations. Along with these, we simulated the respective residual wavefronts and analyzed them for the decorrelation time scale. This analysis suggests that the controlled wavefronts decorrelate slowlier and that the controlled PSFs smear slowlier, allowing at least a factor of $2-3$ longer DITs, depending on the given wavefront error budget and AO mode.

We used the PSFs to create synthetic observations and analyze them for the achievable SNR. We find that longer integration times increase the SNR of the longer spatial wavelengths, as expected, what counteracts the (slower) smearing of the PSF and also allows for longer integration times. This effect is especially prominent at the long spatial wavelength regime, where the SNR increases by a factor of up to 3. Furthermore, applying the GLAO correction is expected to yield a higher SNR in the PSF estimate when using $\Delta H \approx 3$\,mag fainter reference stars, compared to the estimate from noAO data.

We test these findings by applying holographic imaging on synthetic observations with DITs of 1.5\,s and longer and confirm that the reconstruction is significantly better as more flux is shifted from the seeing halo towards the diffraction limited peak. However, in this paper, we concentrate on simulating the turbulence residuals after fast AO correction only. These systems are typically not very robust against very slow ($>1$\,s) opto-mechanical drifts, which in a real system will limit the SNR at high spatial frequencies (and hence the achievable angular resolution) at very long integration times. Our simulation results in Fig.~\ref{fig:power_spectra_integration_time_comparison} suggest, however, that for GLAO corrected NIR imaging, it is worth to check experimentally for a given system and atmosphere the high-resolution coherence time up to the second-long timescales.

The comparison of the holographic imaging technique with the SSA algorithm on VLT/HAWK-I data, obtained with the GLAO correction of the VLT/AOF GRAAL system, as presented in Sect.~\ref{sec:holography__vs_ssa}, clearly suggests the use of the holography technique to remove the residual wavefront aberrations for obtaining diffraction limited imaging. Still, this analysis points out the requirement of a large number of some hundred frames for the technique to beat down the noise in the denominator in Eq.~\ref{eq:reconstructionformula}, where such low frame numbers presented here result in a higher detection limit, compared to the conventional SSA technique.

As a next step, we therefore aim at verifying our results on real observations with fast imaging instruments that are supported by a simple but full-field (GL)AO system, for instance with the combination of LBT/LUCI with the ARGOS LGS GLAO system.
The presented work already strongly suggests the implementation of GLAO assisted imagers with short-exposure imaging modes.

\section*{Acknowledgements}
\acknowledgments{
FB acknowledges support by the Optical Infrared Coordination Network for Astronomy (OPTICON) under a Horizon2020 Grant and from the International Max Planck Research School for Astronomy and Cosmic Physics at the University of Heidelberg (IMPRS-HD). The research leading to these results has received funding from the European Research Council under the European Union's Seventh Framework Programme (FP7/2007-2013) / ERC grant agreement n$^\circ$ [614922].
We thank Marcos van Dam for the help with the YAO software. We appreciate the productive discussions with Richard Mathar, Jochen Heidt, and with Martin Gl{\"u}ck and Vincent Garrel during the setup of the YAO simulations.}

\bibliography{astronomy}

\end{document}